# PoMSA: An Efficient and Precise Position-based Multiple Sequence Alignment Technique

*Sara Shehab*[1,a], *Sameh Shohdy*[1,b], *Arabi E. Keshk*[1,c]

[1]Department of Computer Science, Faculty of Computers and Information, Menoufia University, Shebeen El-Kom 32511, Egypt.
[a]sara.shehab@ci.menofia.edu.eg, [b]sameh.abdulah@ci.menofia.edu.eg, [c]arabi.keshk@ci.menofia.edu.eg

**Abstract—** *Analyzing the relation between a set of biological sequences can help to identify and understand the evolutionary history of these sequences and the functional relations among them. Multiple Sequence Alignment (MSA) is the main obstacle to proper design and develop homology and evolutionary modeling applications since these kinds of applications require an effective MSA technique with high accuracy. This work proposes a novel Position-based Multiple Sequence Alignment (PoMSA) technique -- which depends on generating a position matrix for a given set of biological sequences. This position matrix can be used to reconstruct the given set of sequences in more aligned format. On the contrary of existing techniques, PoMSA uses position matrix instead of distance matrix to correctly adding gaps in sequences which improve the efficiency of the alignment operation. We have evaluated the proposed technique with different datasets benchmarks such as BAliBASE, OXBench, and SMART. The experiments show that PoMSA technique satisfies higher alignment score compared to existing state-of-art algorithms: Clustal-Omega, MAFTT, and MUSCLE.*

## 1.0 INTRODUCTION

The recent huge growth of Bioinformatics usage in different applications requires the development of novel and efficient algorithms that can meet today and future needs in this area. Multiple Sequence Alignment (MSA) is an essential tool for a wide range of these applications where more than two homologous nucleic acid or amino acid sequences need to be aligned to maximize the overall similarity between those sequences.

Practically, the alignment process involves adding one or more gap(s) in different places for each sequence till getting the highest possible matching between the whole sequences set. In recent years, a large amount of research has been devoted to addressing the aforementioned challenge through various techniques, methods, and algorithms: for example, progressive algorithms [1]-[3], Iterative methods [4]-[5], dynamic programming techniques [6], Genetic algorithms [7]-[8], Consensus methods [9]-[10]; see [11] for a review.

In this paper, we propose a Position-based Multiple Sequence Alignment (PoMSA) heuristic algorithm for precise align a set of DNA sequences. The proposed algorithm depends on building a position matrix which helps to effectively add gaps into the given sequences. We have also enhanced PoMSA algorithm by proposing a partitioning scheme that improves the overall performance of the alignment operation.

Further, this work shows a set of extensive experiments to evaluate the proposed algorithm effectiveness. The experiments include a comparison with The state-of-Art algorithms such as Clustal-Omega, MAFFT, and MUSCLE. The experiments show that PoMSA algorithm provides higher matching score compared to these algorithms. Moreover, the experiments show the impact of applying different optimizations to our algorithm (i.e., Different thresholds and partitioning).

## 2.0 RELATED WORK

In literature, several algorithms have been presented as solutions for MSA problem. These algorithms have been divided by the methodology each algorithm uses for solving such a problem. For instance, Carillo and Lipman have proposed a Carillo- Lipman MSA algorithm which is based on Dynamic Programming (DP) [1]. Generally, DP algorithms aim at dividing the whole problem into sub-problems and separately solve each sub-problem with the aid of pre-solved sub-problems. In Carillo-Lipman algorithm, the MSA problem is divided into several pair-wise alignment sub-problems where the main goal is to find the global alignment solution to the whole problem space.

Most the state-of-art MSA algorithms are based on a heuristic search method called progressive alignment. The main goal of these algorithms is to use a clustering method to build a guide tree by adding target sequences to this tree in a sequential way that construct MSA for the given unaligned sequences [15]. For instance, Clustal-Omega depends on guided tree and Hidden Markov Model (HMM) profile-profile technique to align a set of more than two sequences. This Clustal-Omega tool has mainly been proposed to align protein sequences. Further, Fast Fourier Transform (MAFFT) is another tool with the same concept as Clustal-Omega. However, it has been proposed for DNA sequences [16]. Both algorithms take less time to produce the aligned set of sequences but it is less accurate compared to several other tools such as T-Coffee [2]. T-Coffee is slower than both Clustal-Omega and MAFFT but it is more accurate.

## 3.0 DESIGN

### 3.1 Position-based Multiple Sequence Alignment (PoMSA)

In this section, we provide an explanation of the proposed PoMSA algorithm in details. Given $S$ is the set of sequences, $S$ is inserted into a matrix $M$ where the number of rows is fixed and equals to the number of sequences $S$. Initially, the number of columns equals to the length of largest sequence length $L$. If a certain sequence length $l_i$ is less than $L$, $n$ gaps are added to the end of this sequence where $n = L-l_i$ (i.e., adjustment step).

PoMSA scans the position matrix column by column. Assuming a pre-determined threshold ε, If the number of a certain sequence base (i.e., adenine (A) , cytosine (C) , guanine (G), thymine (T)) per column $c_i$ equals or larger than $ε$, this means that we need to set this base as the dominant base for this column. The sequences that have another base in this column should move the unmatched based in a way that increases the overall matching in this column. For unmatched cases, if the previous column $c_{j-1}$ contains a base that equals to current column $c_j$ dominant base, a single gap should be inserted into column $c_{j-1}$ and next bases should be shifted one step forward for this sequence. To keep sequences adjusted a single gap should be added to the end of unchanged sequences. If the next column $c_{j+1}$ have a dominant base, a single gap is added at the end of such a sequence and a single gap should also be added at column $c_i$ for all other sequences.

Figure 1 shows a detailed alignment example using our proposed PoMSA algorithm. In the given examples, PoMSA algorithm requires four steps to get the aligned sequences. Figure (2-a) shows the original set of sequences where each base is represented by a certain character and color. The bases included in the current alignment step are represented by the gray color. In figure (2-a), we start with $col_2$ where we have a dominant base (A). Because the fourth sequence has (G) base in $col_2$, PoMSA algorithm adds a single gap to shift (A) from $col_1$ to $col_2$ as shown by figure (2-b). In figure (2-b), $col_5$ has a dominant base (G). However, the second sequence has (A) at the same column and (G) at $col_6$. Thus, a single gap is added to all sequences – except the second one – so that all of the sequences will have (G) at $col_6$ as shown by figure (2-c). In Figure (2-c), PoMSA algorithm stops at $col_7$ with a new dominant base (G). The first sequence has (A) at $col_7$ and (G) at $col_6$. So, a single gap is added at $col_6$ in the first sequence as shown by figure (2-d). Finally, PoMSA algorithm also finds a dominant (A) at $col_8$. The third sequence has (G) base at this column and (A) base at $col_9$. Thus, a single gap needs to be added to all other sequences at $col_8$ so that all sequences will have the base (A) at $col_9$.

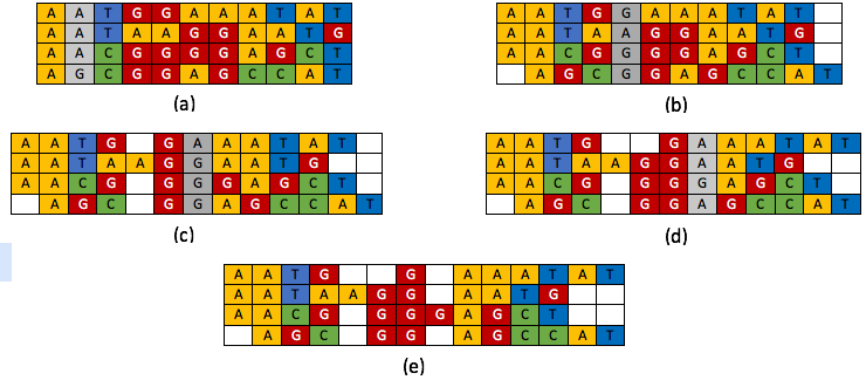

**Figure 1:** PoMSA algorithm example.

Algorithm 1 shows a pseudo code of our proposed PoMSA algorithm. $S$ represents a set of sequences (line 1). In lines 2 and 3, a position matrix $M_{ij}$ is generated and adjusted so that row length is equal to the largest sequence length. Starting from column $col = 0$, the dominant base in each column $col$ is retrieved using *getDominant* function (line 6). For each sequence row, if the base at $col$ does not equal dominant, two more conditions should be checked. First, if the previous base $j-1$ equals dominant, a new gap is added to position $col-1$ and the next bases should be shifted one place to the right for sequence *row* (line 9). Moreover, to keep all sequences justified, a single gap is added to the end of all other sequences (line 10). Second, if the base at $col+1$ matches dominant, a single gap is added to the end of current sequence *row*. For justification, a single gap is added at position col for all other sequences.

**Algorithm 1 : PoMSA Algorithm**
1: Input: $S \rightarrow$ set of sequence.
2: Build a position matrix $M$.
3: Adjust matrix $M$ by adding $L - l_i$ gaps.
4: **for** col:0 to $L$ **do**
5:     **for** row:0 to $S.length$ **do**
6:         dominant=getDominant($M$,col)
7:         **if** M[row][col] != dominant **then**
8:             **if** M[row][col-1] == dominant **then**
9:                 add gap at M[row][col-1].
10:                 add gap to the end of all other sequences.
11:             **end if**
12:             **if** S[row][col+1] == dominant **then**
13:                 add gap at the end of sequence M[row].
14:                 add gap at $col$ for all other sequences.
15:             **end if**
16:         **end if**
17:     **end for**
18: **end for**

**Algorithm 2 : Partitioning Algorithm**
1: Input: $S \rightarrow$ set of sequence.
2: Let k=0;
3: **for** col:0 to $L$ **do**
4:     **for** row:0 to $S.length$ **do**
5:         **if** $S[col]$ is the same for all $S$ **then**
6:             alignedPartition=pomsa($S[k] \Rightarrow S[j]$)
7:             k=j+1
8:             alighnedSeqs+=alignedPartition
9:         **end if**
10:     **end for**
11: **end for**

### 3.2 PoMSA Algorithm Partitioning Scheme

We also propose a partitioning scheme that increases the total alignment score of the given set of sequences if combined with PoMSA algorithm. Simply, we partition the sequences into *P* partitions. The splitting points are the *j* locations where all the sequences have the same base. After this, each partition is separately processed by the PoMSA algorithm. Algorithm 2 shows a pseudo code for partitioning operation. The algorithm has the set of sequences *S* as input. If all sequences have the same base at position *col* (line 5), the algorithm splits the sequences at this point and call the main PoMSA algorithm (i.e., Algorithm 1) starting from the last split point to this point. In this case, for *P* partitions, algorithm 2 calls PoMSA algorithm *P* times. Finally, all the results retrieved from algorithm 1 should be merged to get the complete set of aligned sequences *alighnedSeqs* (line 8).

### 4.0 EXPERIMENTAL RESULTS

In this section, we evaluate the performance of the proposed PoMSA algorithm by providing three sets of experiments with three different goals. First, we evaluate the impact of using partitioning with the PoMSA algorithm. Second, PoMSA algorithm is evaluated using different matching threshold ε. Third, PoMSA algorithm is compared against three different common MSA algorithms: Clustal-Omega [11], MAFTT [1], and MUSCLE [4]. To provide a fair comparison, the experiments have been applied to three different datasets: BAliBASE [13], SMART [5], and OXBench [14]. Moreover, all the experiments use Sum-of-Pairs (SP) score to measure the alignment efficiency for different algorithms [1].

### 4.1 Impact of Using Partitioning with PoMSA

In this experiment, we compare two different versions of our proposed PoMSA algorithm: PoMSA w partitioning and PoMSA w/o partitioning. Here, we aim at studying the impact of the proposed partitioning scheme on PoMSA algorithm. The experiment has been conducted using different sequences files from each given dataset. Figure 2 shows the performance of both PoMSA versions according to the overall alignment score (SP). The x-axis represents the sequences files and the y-axis represents the final alignment score. As shown, applying partitioning to PoMSA algorithm improves the overall alignment score with different sequences datasets. Precisely, using BAliBASE, OXBench, and SMART datasets, the average improvement is about 4.31x, 0.65x, and 2.39x, respectively.

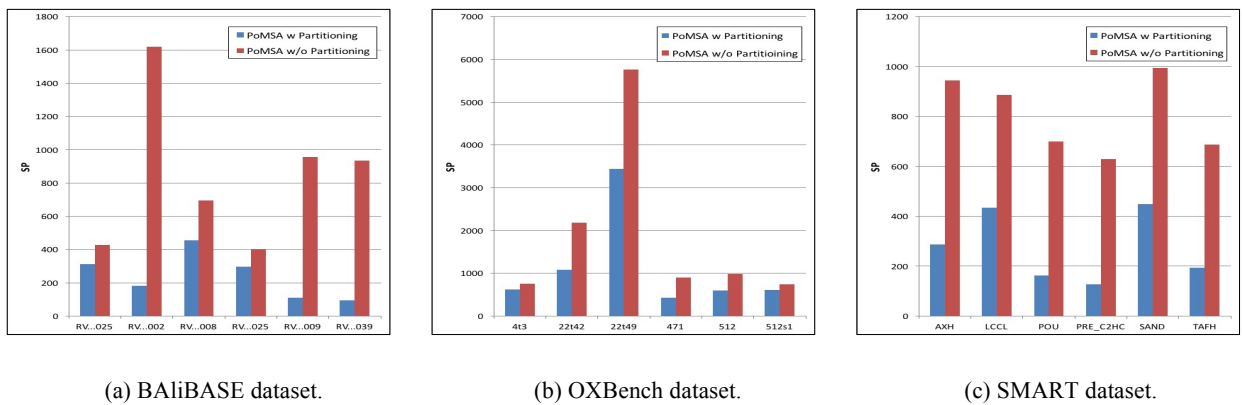

(a) BAliBASE dataset.　　　　(b) OXBench dataset.　　　　(c) SMART dataset.

**Figure 2:** Impact of using partitioning with PoMSA algorithm.

### 4.2 Impact of Different Thresholds ε

Gaps insertion decisions depend on the number of matched bases through the whole set of sequences. When the matching percentage exceeds a pre-determined threshold ε, one or more gap(s) could be added to the target sequences at different positions. In this experiment, we evaluate the using of three different matching threshold ε - i.e., 0.75, 0.85 and 0.95. As we aim to provide an optimum version of our proposed PoMSA algorithm, we have performed this experiment using PoMSA with partitioning version.

Figure 3 shows the impact of three different thresholds ε to PoMSA algorithm on BAliBASE, OXBench, and SMART datasets. As shown, 0.75 thresholds satisfies the highest alignment score compared to 0.85 and 0.95 thresholds. Using BAliBASE dataset, 0.75 threshold satisfies average improvement of 0.58x, and 0.07x compared to 0.85, and 0.95, respectively. Using OXBench dataset, the average improvement is 0.37x with ε=0.85, and 0.01x with ε=0.95. Finally, using SMART dataset, the average improvement is 0.71x with ε=0.85, and 0.06x with ε=0.95.

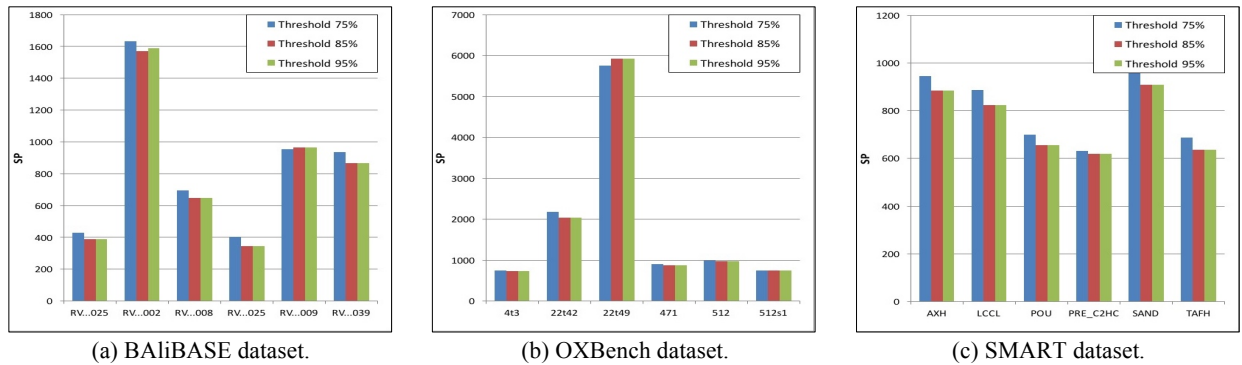

(a) BAliBASE dataset.    (b) OXBench dataset.    (c) SMART dataset.

**Figure 3:** Three different thresholds ε – 0.75, 0.85 and 0.95 – impact over PoMSA algorithm.

### 4.3 PoMSA Evaluation Compared to the State-of-Art Algorithms

Even partitioning and tuning the threshold value of PoMSA algorithm shows a tangible improvement, optimum version of PoMSA algorithm should be compared with the existing state-of-art algorithms. Mainly, we have used three different MSA algorithms (i.e., Clustal-Omega, MAFFT, and MUSCLE). Figure 4 shows the performance over target datasets. As shown, Clustal-Omega, MAFFT and MUSCLE algorithms have disparate performance over different sets of sequences. However, our proposed PoMSA algorithm has better performance overall given sets.

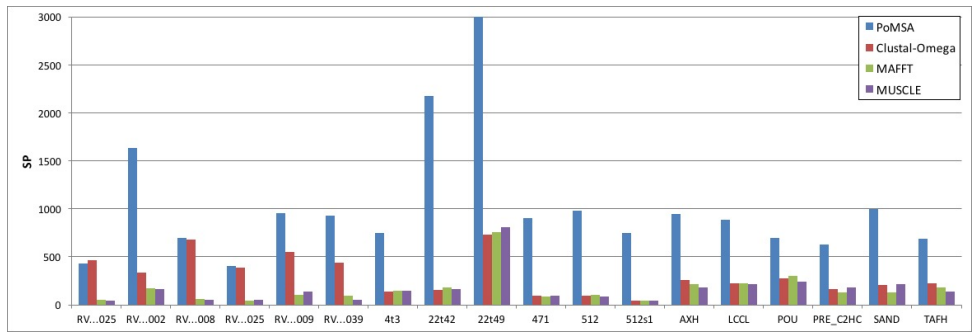

**Figure 4:** Comparing PoMSA with Clustal-Omega, MAFFT, and MUSCLE algorithms.

## 5.0 CONCLUSION

In this paper, we propose a novel Multiple Sequence Alignment (MSA) algorithm, i.e., Position-based Multiple Sequence Alignment (PoMSA). This algorithm depends on building a position matrix that can be efficiently used to better align a set of sequences. We support our algorithm by providing a partitioning scheme so that each partition can separately be processed by the algorithm and satisfies higher matching score. Moreover, we have compared the PoMSA algorithm with a set of existing algorithms such as Clustal-Omega, MAFFT, and MUSCLE and showed that PoMSA is able to satisfy higher matching score compared to these algorithms.